\documentclass{article}
\usepackage{times}
\usepackage{epsfig,amssymb,amsfonts,amsmath}

\def\eeq{\end{equation}}
\def\beq{\begin{equation}}
\def\bea{\begin{eqnarray}}
\def\eea{\end{eqnarray}}

\providecommand{\rbr}[1]{\left( #1 \right)}%
\providecommand{\sqbr}[1]{\left[ #1 \right]} %

\begin{document}

\title{Comment on "Essential discreteness in generalized thermostatistics with non-logarithmic entropy"
by S. Abe}

\author{G. Baris Bagci$^{1,}$\thanks{Corresponding author: baris.bagci@ege.edu.tr} ,
Thomas Oikonomou$^{2}$, Ugur Tirnakli$^{1}$ \\ \\
$^1$ Department of Physics, Faculty of Science, \\
Ege University, 35100 Izmir, Turkey \\ \\
$^2$ Institute of Physical Chemistry, \\
National Center for Scientific Research ``Demokritos", \\
15310 Athens, Greece\\
}

\date{\today}

\maketitle

\begin{abstract}
Recently Abe (arXiv:cond-mat/1005.5110v1) claimed that
the $q$-entropy of nonextensive statistical mechanics cannot be generalized for 
the continuous variables
and therefore can be used \textit{only} in the discrete case.
In this letter, we show that the discrete $q$-entropy can be generalized
to continuous variables \textit{exactly in the same manner as}
Boltzmann-Gibbs entropy, contrary to the claim by Abe, so that $q$-entropy
can be used with discrete as well as continuous variables.\\

PACS: 05.20.-y, 05.70.-a

\end{abstract}

\newpage

\section{Introduction}\label{intro}

As Abe very recently reminded \cite{Abecomment},
Boltzmann-Gibbs (BG) entropy has originated mainly for the case
when the random variable takes finite number of values \cite{Uffink}.
When this is the case, BG entropy, setting Boltzmann constant to unity,
is given by

\begin{equation}\label{BG}
S(p_{i})=-\sum\limits_{i=1}^{n}p_{i}\ln p_{i}.
\end{equation}
In order to generalize the discrete expression above to the
continuous case, the usual practice is to assume the existence of
a probability density function $p(x)$, which is equal to or
greater than zero in some interval $[a,b]$. In addition to this,
the criterion of normalization is assumed to be satisfied by
$p(x)$ in the interval $[a,b]$. Then, in a straightforward manner,
Eq.~\eqref{BG} is extended to the continuous variables as

\begin{equation}\label{BGint}
S(p)=-\int\limits_{a}^{b}p(x)\ln p(x)dx.
\end{equation}

Although this generalization by itself seems reasonable, three
serious drawbacks can be noted at once. First, probability
densities usually have a unit of inverse of length in one
dimensional continuum. Then, $S(p)$ possesses an overall unit of
$\log($length$)$ in contrast to the dimensionless discrete entropy
expression $S(p_{i})$ in Eq.~\eqref{BG}. Second, $S(p)$ is not
invariant with respect to coordinate transformations deeming it
vulnerable to Bertrand paradox \cite{Uffink}. The last but not the
least, the discrete BG entropy $S(p_{i})$ in the
$n\rightarrow\infty$ limit and $S(p)$ yield different results. In
order to show this, we consider a uniform distribution in the
interval $[a,b]$ i.e.,

\begin{equation}\label{ornek1}
p(x)=\frac{1}{b-a}
\end{equation}
such that the corresponding probabilities in the discrete case,
assuming that the interval $[a,b]$ is divided in $n$ equal
subintervals, are of the form

\begin{equation}\label{ornek2}
p(x_{i})=\frac{1}{n}
\end{equation}
where $i=1,2,...,n$. The calculation of the continuous entropy
$S(p)$ with the probability density given by Eq.~\eqref{ornek1}
results in

\begin{equation}\label{ornek3}
S(p)=\ln(b-a)
\end{equation}
whereas the discrete form $S(p_{i})$ together with Eq.~\eqref{ornek2}
yield

\begin{equation}\label{ornek3}
S(p_{i})=\ln(n).
\end{equation}
The discrete entropy $S(p_{i})$ attains the infinity in the
$n\rightarrow\infty$ limit, whereas the continuous version $S(p)$
is independent of $n$, and again equal to $\ln(b-a)$. Summing up,
the so-called continuous version $S(p)$ of the discrete BG entropy
$S(p_{i})$ is not tenable and cannot be used in the continuum.

Although the solution to this dilemma is known in the case of BG
entropy in detail \cite{Uffink} as reviewed in the next section
too for the sake of completeness of our treatment, Abe
\cite{Abecomment} recently argued that the same cannot be said for
the $q$-entropy proposed by Tsallis \cite{Tsallis88, Tsallisbook}
and deduced that the $q$-entropy is properly functional only in
the discrete case. In this paper, we show that the $q$-entropy can
indeed be generalized to the continuous case as naturally as BG entropy
so that Abe's desideratum is satisfied and therefore the conclusion
reached by Abe is shown to be unfounded.

\section{Extension to the Continuum: Boltzmann-Gibbs entropy}\label{Main Idea}

In this section we review the generalization of BG entropy in the
case of random variables, since $q$-entropy is generalized almost
in the same manner as BG entropy. Despite the fact that one can
have different derivations of this generalization depending on the
level of rigor, we follow Abe's treatment \cite{Abecomment}, which
is identical to the one by Jaynes (see p. 375 in Ref.~\cite{Jaynes} for example). 
In order to extend BG entropy to the
continuum, we consider an interval $[a,b]$ so that the discrete
points $x_{i}$ with $i=1,2,...,n$ fill the interval. Then, the
relation between the discrete probability $p_{i}$ and the
probability density $\rho(x_{i})$ is given by \cite{Abecomment,
Jaynes}

\begin{equation}\label{interval}
p_{i}\rightarrow \frac{\rho \left( x_{i}\right) }{nm\left(
x_{i}\right) }.
\end{equation}
As the number of points increases and tends to infinity, one can
write \cite{Abecomment, Jaynes}

\begin{equation}\label{continuum}
\lim_{n\rightarrow \infty }\sum\limits_{i=1}^{n}\frac{1}{nm\left(
x_{i}\right) }=\int\limits_{a}^{b}dx.
\end{equation}
Substitution of Eq.~\eqref{interval} into the discrete entropy
expression given by Eq.~\eqref{BG} yields

\begin{equation}\label{BGint1}
S=-\sum\limits_{i=1}^{n}\frac{\rho \left( x_{i}\right) }{nm\left(
x_{i}\right) }\ln \left( \frac{\rho \left( x_{i}\right) }{nm\left(
x_{i}\right) }\right).
\end{equation}
Eq.~\eqref{BGint1} can now be rewritten as

\begin{equation}\label{BGint2}
S=-\sum\limits_{i=1}^{n}\frac{\rho \left( x_{i}\right) }{nm\left(
x_{i}\right) }\ln \left( \frac{\rho \left( x_{i}\right) }{m\left(
x_{i}\right) }\right) +\sum\limits_{i=1}^{n}\frac{\rho \left(
x_{i}\right) }{nm\left( x_{i}\right) }\ln \left( n\right)
\end{equation}
so that Eq.~\eqref{BGint2}, using Eq.~\eqref{continuum}
in the $n\rightarrow\infty$ limit,
becomes

\begin{equation}\label{relent1}
S=-\int\limits_{a}^{b}dx\rho \left( x\right) \ln \left( \frac{\rho
\left( x\right) }{m\left( x\right) }\right) +\ln \left( n\right)
\int\limits_{a}^{b}dx\rho \left( x\right).
\end{equation}
It is crucial to understand that the left hand side of the
equation above is discrete BG entropy given by Eq.~\eqref{BG}, but
when one considers it in the $n\rightarrow\infty$ limit i.e.,
continuous limit. Eq.~\eqref{relent1}, due to normalization, can
be rewritten as

\begin{equation}\label{relent2}
S=-\int\limits_{a}^{b}dx\rho \left( x\right) \ln \left( \frac{\rho
\left( x\right) }{m\left( x\right) }\right) \, 
\end{equation}
where the additive divergent term $\lim_{n\rightarrow\infty}\ln \left( n\right)$ 
is omitted since the entropy is not absolute but only its change can be
measured. 
Then, this equation indicates that the
continuous form of BG entropy (remember that the left hand side of
the equation above is discrete BG entropy but in the
$n\rightarrow\infty$ limit) is given by the negative of the relative entropy i.e.,
right hand side of Eq.~\eqref{relent2} as discussed in \cite{Abecomment, Jaynes}.
The integral in the right hand side i.e.,
the extension of BG entropy to continuum is called relative
entropy or Kullback-Leibler entropy, since it was first proposed
by Kullback and Leibler \cite{KL}. 
It is worth remarking that the relative entropy expression given
by Eq.~\eqref{relent2} is dimensionless like its discrete
counterpart, and invariant under different reparametrization of
continuum.

\section{Extension to the Continuum: $q$-entropy}\label{original}

BG entropy is the cornerstone of statistical mechanics and
its maximization yields exponential distributions. On the other
hand, there are many physical systems exhibiting inverse power law
distributions (see Ref.~\cite{Tsallisbook} for a survey).
Therefore, the discrete $q$-entropy was proposed
\cite{Tsallis88, Tsallisbook} as a generalization of BG entropy
measure in order to investigate certain classes of such systems.
The nonadditive $q$-entropy reads

\begin{equation}\label{q-entropy1}
S_{q}(p_{i})=\sum\limits_{i=1}^{n}p_{i}\ln
_{q}\left(\frac{1}{p_{i}}\right)=-\sum_{i=1}^n p_i\ln_{2-q}(p_i)
\end{equation}
where $q$-logarithm is defined as
\begin{equation}\label{q-log}
\ln _{q}(x)=\frac{x^{1-q}-1}{1-q}
\end{equation}
for $x>0$. Note that $q$-logarithm becomes the ordinary
logarithmic function in the $q\rightarrow1$ limit so that the
discrete $q$-entropy in Eq.~\eqref{q-entropy1} takes the form of
ordinary BG entropy.

In order to extend the discrete $q$-entropy definition in
Eq.~\eqref{q-entropy1} to continuum, we consider an interval $[a,b]$
filled by the discrete points $x_{i}$ where the index $i$ runs
up to $n$, exactly as in the previous section. Then, the relation
between the discrete probability $p_{i}$ and the probability
density $\rho(x_{i})$ \cite{Abecomment, Jaynes} is given by Eq.~\eqref{interval} 
with $m\rightarrow m_q$, where $m_q(x_i)$ is the $q$-generalized $m(x_i)$ measure, 
relevant for a nonextensive system, as mentioned also in Ref.~\cite{Abecomment}, 
and present the essential difference between ordinary and $q$-statistics.
Therefore, one can write

\begin{equation}\label{q-entropy2}
S_{q}=-\sum\limits_{i=1}^{n}\frac{\rho \left( x_{i}\right)
}{n\,m_q\left( x_{i}\right) }\ln _{2-q}\left( \frac{\rho
\left( x_{i}\right) }{n\,m_q\left( x_{i}\right) }\right)\, .
\end{equation}
Let us now define the $m_q(x_i)$ measure as follows
\begin{equation}\label{mq-measure}
m_q(x_i):=\frac{\rho(x_i)}{n}\sqbr{\rbr{\frac{\rho(x_i)}{m(x_i)}}\oslash_{2-q}n}^{-1}
\end{equation}
where $\oslash_{2-q}$ is the generalized division originally introduced 
in \cite{Borges,Wang} as $x\oslash_{q} y = \left[x^{1-q}-y^{1-q}+1\right]^{1/(1-q)}$, 
for $q\rightarrow2-q$  and $m_{q\rightarrow1}(x_i)=m(x_i)$. 
The measure $m_q(x_i)$ is consistently defined since it has again the dimension of 
a probability and let $\Delta x_i$ tend to zero for $n\rightarrow\infty$.
Then, by virtue of Eq.~\eqref{mq-measure}, we obtain from Eq.~\eqref{q-entropy2}
\begin{equation}\label{q-entropy5}
S_{q}=-\sum_{i=1}^{n}\frac{\rho \left( x_{i}\right)}{n\,m_q\left( x_{i}\right) }
\ln _{2-q}\rbr{\frac{\rho(x_i)}{m(x_i)}}+\ln_{2-q}(n)\sum_{i=1}^{n}\frac{\rho \left( x_{i}\right)}{n\,m_q\left( x_{i}\right) } \, .
\end{equation}
Using Eq.~\eqref{continuum} in the $n\rightarrow\infty$ limit with $m\rightarrow m_q$, we
can write Eq.~\eqref{q-entropy5} in the continuous case as
\begin{equation}\label{q-entropy6}
S_{q}=- \int\limits_{a}^{b}dx\rho(x)\ln_{2-q}\rbr{\frac{\rho(x)}{m(x)}}+
\ln_{2-q}(n)\int\limits_{a}^{b}dx\rho \left( x\right)\,.
\end{equation}
It is clearly seen that the left hand side of this equation is discrete $q$-entropy 
given by Eq.~\eqref{q-entropy1} when it is considered in the $n\rightarrow\infty$ limit. 
Eq.~\eqref{q-entropy6}, due to normalization, can be rewritten as

\begin{equation}\label{q-entropy7}
S_{q}=- \int\limits_{a}^{b}dx\rho(x)\ln_{2-q}\rbr{\frac{\rho(x)}{m(x)}} \, ,
\end{equation}
since the additive divergent term  $\lim_{n\rightarrow\infty}\ln_{2-q} \left( n\right)$ 
is omitted due to the fact that the entropy is not absolute but only its change 
can be measured. 
This result can also be obtained by plugging the definition in Eq.~\eqref{mq-measure} 
into Eq.~(10) of Ref.~\cite{Abecomment}.
This last equation i.e., Eq.~\eqref{q-entropy6}, is essential,
since it shows that the continuous form of $q$-entropy (just like
BG case in the previous section, note that the left hand side of
the equation above is discrete $q$-entropy but in the
$n\rightarrow\infty$ limit) is given by the negative of the corresponding
$q$-relative entropy i.e., the integral in the right hand side of
Eq.~\eqref{q-entropy7} \cite{AbeBagci}.

It can now be seen that none of the divergences mentioned by Abe
\cite{Abecomment} occurs in Eq.~\eqref{q-entropy7}, since the
divergence is, again like BG entropy, additive and therefore can be omitted
(compare Eq.~(10) in Ref.~\cite{Abecomment} and Eq.~\eqref{q-entropy7}
in the present manuscript). Moreover, the
relative entropy expression given by Eq.~\eqref{q-entropy7} is
dimensionless and invariant under
different reparametrization of continuum just like its BG
counterpart.

\section{Conclusions}
Contrary to the recent claim by Abe \cite{Abecomment}, we have
shown that the discrete $q$-entropy can be generalized to the
continuous variables exactly in the same manner as done for
the BG entropy. Moreover, the resulting generalization of
the $q$-entropy to the continuum is dimensionless and independent
of reparametrization of the continuous variables just as the
continuous BG entropy.

\section*{Acknowlegment}
GBB was supported by Faculty of Science at Ege
University under the project number 2009Fen076.


\end{document}